 %
 %
\input harvmac.tex  
 %
\catcode`@=11
\def\rlx{\relax\leavevmode}                  
 %
 %
 %
\font\tenmib=cmmib10
\font\sevenmib=cmmib10 at 7pt 
\font\fivemib=cmmib10 at 5pt  
\font\tenbsy=cmbsy10
\font\sevenbsy=cmbsy10 at 7pt 
\font\fivebsy=cmbsy10 at 5pt  
\def\BMfont{\textfont0\tenbf \scriptfont0\sevenbf
                              \scriptscriptfont0\fivebf
            \textfont1\tenmib \scriptfont1\sevenmib
                               \scriptscriptfont1\fivemib
            \textfont2\tenbsy \scriptfont2\sevenbsy
                               \scriptscriptfont2\fivebsy}
\def\BM#1{\rlx\ifmmode\mathchoice
                      {\hbox{$\BMfont#1$}}
                      {\hbox{$\BMfont#1$}}
                      {\hbox{$\scriptstyle\BMfont#1$}}
                      {\hbox{$\scriptscriptstyle\BMfont#1$}}
                 \else{$\BMfont#1$}\fi}
 %
 %
 %
 %
\def\inbar{\vrule height1.5ex width.4pt depth0pt}
\def\sinbar{\vrule height1ex width.35pt depth0pt}
\def\ssinbar{\vrule height.7ex width.3pt depth0pt}
\font\cmss=cmss10
\font\cmsss=cmss10 at 7pt
\def\ZZ{\rlx\leavevmode
             \ifmmode\mathchoice
                    {\hbox{\cmss Z\kern-.4em Z}}
                    {\hbox{\cmss Z\kern-.4em Z}}
                    {\lower.9pt\hbox{\cmsss Z\kern-.36em Z}}
                    {\lower1.2pt\hbox{\cmsss Z\kern-.36em Z}}
               \else{\cmss Z\kern-.4em Z}\fi}
\def\Ik{\rlx{\rm I\kern-.18em k}}  
\def\IC{\rlx\leavevmode
             \ifmmode\mathchoice
                    {\hbox{\kern.33em\inbar\kern-.3em{\rm C}}}
                    {\hbox{\kern.33em\inbar\kern-.3em{\rm C}}}
                    {\hbox{\kern.28em\sinbar\kern-.25em{\sevenrm C}}}
                    {\hbox{\kern.25em\ssinbar\kern-.22em{\fiverm C}}}
             \else{\hbox{\kern.3em\inbar\kern-.3em{\rm C}}}\fi}
\def\IP{\rlx{\rm I\kern-.18em P}}
\def\IR{\rlx{\rm I\kern-.18em R}}
\def\Ione{\rlx{\rm 1\kern-2.7pt l}}
 %
 %

 %

\def\intem#1{\par\leavevmode%
              \llap{\hbox to\parindent{\hss{#1}\hfill~}}\ignorespaces}
 %


 %
\newskip\humongous \humongous=0pt plus 1000pt minus 1000pt   
\def\caja{\mathsurround=0pt}
\newif\ifdtup
 %
\def\eqalign#1{\,\vcenter{\openup2\jot \caja
     \ialign{\strut \hfil$\displaystyle{##}$&$
      \displaystyle{{}##}$\hfil\crcr#1\crcr}}\,}
 %

 %
\def\panorama{\global\dtuptrue \openup2\jot \caja
     \everycr{\noalign{\ifdtup \global\dtupfalse
      \vskip-\lineskiplimit \vskip\normallineskiplimit
      \else \penalty\interdisplaylinepenalty \fi}}}
 %

 %
\def\eqalignnotwo#1{\panorama \tabskip=\humongous
     \halign to\displaywidth{\hfil$\displaystyle{##}$
      \tabskip=0pt&$\displaystyle{{}##}$
       \tabskip=0pt&$\displaystyle{{}##}$\hfil
        \tabskip=\humongous&\llap{$##$}\tabskip=0pt\crcr#1\crcr}}
 %

 %

 %
 %
 %
 %
   \let\SS=\S       
\def\,{\hskip1.5pt}           
 %

\let\c=\chi
\let\d=\delta       \let\vd=\partial             
\let\e=\epsilon

\let\h=\eta

\let\j=\psi                                      

                                   \let\L=\Lambda

\let\s=\sigma                   \let\S=\Sigma

\let\w=\omega

 %
 %
\def\Box{\sqcap\llap{$\sqcup$}}
\def\lapp{\lower.4ex\hbox{\rlap{$\sim$}} \raise.4ex\hbox{$<$}}
\def\gapp{\lower.4ex\hbox{\rlap{$\sim$}} \raise.4ex\hbox{$>$}}
\def\con{\ifmmode\raise.1ex\hbox{\bf*}
          \else\raise.1ex\hbox{\bf*}\fi}
\def\bo{{\raise.15ex\hbox{\large$\Box\kern-.39em$}}}
\let\iff=\leftrightarrow

\def\dual{\relax\leavevmode\lower.9ex\hbox{\titlerms*}}
\def\define{\buildrel\rm def\over =}

\let\8=\otimes
 %
 %
 %
 %

\let\2=\underline

 %
\def\dt#1{{\buildrel{\smash{\lower1pt\hbox{.}}}\over{#1}}}

\font\eightrm=cmr8
\def\6(#1){\relax\leavevmode\hbox{\eightrm(}#1\hbox{\eightrm)}}
\def\0#1{\relax\ifmmode\mathaccent"7017{#1}     
                \else\accent23#1\relax\fi}      
\def\7#1#2{{\mathop{\null#2}\limits^{#1}}}      
\def\5#1#2{{\mathop{\null#2}\limits_{#1}}}      
 %

\def\V#1{\langle#1\rangle}

 %

 %

 %

\def\too#1{\buildrel{#1}\over{\to}}
 %
\newbox\t@b@x
\def\rightarrowfill{$\m@th \mathord- \mkern-6mu
     \cleaders\hbox{$\mkern-2mu \mathord- \mkern-2mu$}\hfill
      \mkern-6mu \mathord\rightarrow$}
\def\tooo#1{\setbox\t@b@x=\hbox{$\scriptstyle#1$}%
             \mathrel{\mathop{\hbox to\wd\t@b@x{\rightarrowfill}}%
              \limits^{#1}}\,}
\def\leftarrowfill{$\m@th \mathord\leftarrow \mkern-6mu
     \cleaders\hbox{$\mkern-2mu \mathord- \mkern-2mu$}\hfill
      \mkern-6mu \mathord-$}
\def\froo#1{\setbox\t@b@x=\hbox{$\scriptstyle#1$}%
             \mathrel{\mathop{\hbox to\wd\t@b@x{\leftarrowfill}}%
              \limits^{#1}}\,}
 %
\def\frac#1#2{{#1\over#2}}
\def\frc#1#2{\relax\ifmmode{\textstyle{#1\over#2}} 
                    \else$#1\over#2$\fi}           
\def\inv#1{\frc{1}{#1}}                            
 %
\def\Claim#1#2#3{\bigskip\begingroup%
                  \xdef #1{\secsym\the\meqno}%
                   \writedef{#1\leftbracket#1}%
                    \global\advance\meqno by1\wrlabeL#1%
                     \noindent{\bf#2}\,#1{}\,:~\sl#3\vskip1mm\endgroup}

\def\QED{\rlx\hfill$\Box$\kern-7pt\raise3pt\hbox{$\surd$}\bigskip}
 %
 %

 %
\def\muthstrut{\vphantom1}
\def\mutrix#1{\null\,\vcenter{\normalbaselines\m@th
        \ialign{\hfil$##$\hfil&&~\hfil$##$\hfill\crcr
            \muthstrut\crcr\noalign{\kern-\baselineskip}
            #1\crcr\muthstrut\crcr\noalign{\kern-\baselineskip}}}\,}

 %
\def\YT#1#2{\vcenter{\hbox{\vbox{\baselineskip0pt\parskip=\medskipamount%
             \def\Box{$\sqcap\llap{$\sqcup$}$\kern-1.2pt}%
              \def\Z{\hfil\vskip-5.8pt}\lineskiplimit0pt\lineskip0pt%
               \setbox0=\hbox{#1}\hsize\wd0\parindent=0pt#2}\,}}}
\def\EU{\rlx\ifmmode \c_{{}_E} \else$\c_{{}_E}$\fi}
\def\TM{\rlx\ifmmode {\cal T_M} \else$\cal T_M$\fi}
\def\TW{\rlx\ifmmode {\cal T_W} \else$\cal T_W$\fi}
\def\CM{\rlx\ifmmode {\cal T\rlap{\bf*}\!\!_M}
             \else$\cal T\rlap{\bf*}\!\!_M$\fi}
\def\hm#1#2{\rlx\ifmmode H^{#1}({\cal M},{#2})
                 \else$H^{#1}({\cal M},{#2})$\fi}
\def\CP#1{\rlx\ifmmode\IP^{#1}\else\IP$^{#1}$\fi}
\def\cP#1{\rlx\ifmmode\IC{\rm P}^{#1}\else$\IC{\rm P}^{#1}$\fi}

\def\sll#1{\rlx\rlap{\,\raise1pt\hbox{/}}{#1}}
\def\Sll#1{\rlx\rlap{\,\kern.6pt\raise1pt\hbox{/}}{#1}\kern-.6pt}
%

 %
 %

\def\CY{Calabi-\kern-.2em Yau}

\def\3{\ifmmode\ldots\else$\ldots$\fi}
\def\Z{\hfil\break\rlx\hbox{}\quad}
\def\3{\ifmmode\ldots\else$\ldots$\fi}
\def\?{d\kern-.3em\raise.64ex\hbox{-}}           
\def\9{\raise.43ex\hbox{-}\kern-.37em D}         

 %
 %
\def\I#1{{\it ibid.\,}{\bf#1\,}}

\def\NP#1{{\it Nucl.\,Phys.\,}{\bf#1\,}}

\def\MPL#1{{\it Mod.\,Phys.\,Lett.\,}{\bf#1\,}}
\def\PRL#1{{\it Phys.\,Rev.\,Lett.\,}{\bf#1\,}}
\def\CMP#1{{\it Commun.\,Math.\,Phys.\,}{\bf#1\,}}

 %
 %
 %
\baselineskip=13.0861pt plus2pt minus1pt
\parskip=\medskipamount
\let\ft=\foot
\noblackbox
\def\SaveTimber{\abovedisplayskip=1.5ex plus.3ex minus.5ex
                \belowdisplayskip=1.5ex plus.3ex minus.5ex
                \abovedisplayshortskip=.2ex plus.2ex minus.4ex
                \belowdisplayshortskip=1.5ex plus.2ex minus.4ex
                \baselineskip=12pt plus1pt minus.5pt
 \parskip=\smallskipamount
 \def\ft##1{\unskip\,\begingroup\footskip9pt plus1pt minus1pt\setbox%
             \strutbox=\hbox{\vrule height6pt depth4.5pt width0pt}%
              \global\advance\ftno by1\footnote{$^{\the\ftno)}$}{##1}%
               \endgroup}
 \def\listrefs{\footatend\vfill\immediate\closeout\rfile%
                \writestoppt\baselineskip=10pt%
                 \centerline{{\bf References}}%
                  \bigskip{\frenchspacing\parindent=20pt\escapechar=` %
                   \rightskip=0pt plus4em\spaceskip=.3333em%
                    \input refs.tmp\vfill\eject}\nonfrenchspacing}}
 %
\def\Afour{\ifx\answ\bigans
            \hsize=16.5truecm\vsize=24.7truecm
             \else
              \hsize=24.7truecm\vsize=16.5truecm
               \fi}
\catcode`@=12
 %
 %
\def\rd{{\rm d}}
\def\\{\hfill\break}
\def\Qb{\relax\leavevmode\hbox{$Q$\kern-.55em\vrule
         height1.9ex depth-1.8ex width5.5pt}\kern.5pt\vphantom{Q}}
\def\db{\relax\leavevmode\hbox{$\partial$\kern-.4em\vrule
         height1.86ex depth-1.8ex width4pt}\kern.8pt\vphantom{\vd}}
 %
 %
 %
\Title{HUPAP-96/101}
      {\vbox{\centerline{On A Stringy Singular Cohomology}}}
 \centerline{Tristan H\"ubsch\footnote{$^{\spadesuit}$}
      {On leave from the Institute ``Ru\?er Bo\v skovi\'c'',
       Zagreb, Croatia.}}                             \vskip0mm
 \centerline{Departments of Physics and Astronomy}    \vskip-1mm
 \centerline{Howard University, Washington, DC~20059} \vskip-1mm
 \centerline{hubsch\,@\,scs.howard.edu}
\vfill

\centerline{ABSTRACT}\vskip2mm
\vbox{\leftskip=5.6em\rightskip=\leftskip\baselineskip=12pt\noindent
String theory has already motivated, suggested, and sometimes well-nigh proved
a number of interesting and sometimes unexpected mathematical results, such as
mirror symmetry. A careful examination of the behavior of string propagation
on (mildly) singular varieties similarly suggests a new type of (co)homology
theory. It has the `good behavior' of the well established intersection
(co)homology and $L^2$-cohomology, but is markedly different in some aspects.
For one, unlike the intersection (co)homology and the $L^2$-cohomology (or any
other known thus far), this new cohomology is symmetric with respect to the
mirror map. Among the available choices, this makes it into a prime candidate
for describing the string theory zero modes in geometrical terms.}

\Date{October 1996}
\footline{\hss\tenrm--\,\folio\,--\hss}
\noblackbox
 %
 %
\lref\rArnold{V.I.~Arnold: in {\it Proceedings of the International
       Congress of Mathematicians}, p.19, Vancouver, 1974, {\it
       Singularity Theory} ({\it London Math.\ Soc.\ Lecture Note
       Series}~53, Cambridge University Press, Cambridge, 1981)\semi
      {V.I.~Arnold, S.M.~Gusein-Zade and A.N.~Varchenko}: {\it
       Singularities of Differentiable Maps} (Birkh\"auser, Boston,
       1985).}

\lref\rAGM{P.S.~Aspinwall,  B.R.~Greene and D.R.~Morrison:
       alg-geom/9309007; \NP{B420}(1994)184--242;
       in the Proceedings ``Strings '93'' (Berkeley 1993), p.241--262;
       hep-th/9311186.}

\lref\rExo{P.S.~Aspinwall,  B.R.~Greene and D.R.~Morrison:
       \NP{B416}(1994)414--480.}

\lref\rAtSe{M.~Atyah and G.~Segal: {\it
       J.\,Geom.\,Phys.\,\bf6}(1989)671--677.}

\lref\rBrGr{R.~Bryant and P.~Griffiths: {\it Progress in Mathematics}
      {\bf 36} pp.77--102 (Birkh\"auser, Boston, 1983).}

\lref\rCCC{P.~Candelas, P.~Green and T.~H\"ubsch:
       \PRL{62}(1989)1956--1959;
       in ``Strings '88'', p.155, eds.~S.J.~Gates~Jr., C.R.~Preitschopf and
       W.~Siegel (World Scientific, Singapore, 1989);
       \NP{B330}(1990)49--102.}

\lref\rIHOM{J.~Cheeger, M.~Goresky:
       {\it Proc.\,Am.\,Math.\,Soc.\,\bf72}(1978)193--200\semi
       M.~Goresky and R.~MacPherson: {\it Topology\,\bf19}(1980)135--162\semi
       M.~Goresky and R.~MacPherson:
       {\it Invent.\,Math.\,\bf71}(1983)77--129.}

\lref\rIHom{J.~Cheeger, M.~Goresky and R.~MacPherson: {\it
       Ann.\,Math.\,Studies~\bf102}(1982)303--340.}

\lref\rDHVW{L.~Dixon, J.~Harvey, C.~Vafa and E.~Witten:
       \NP{B261}(1985)678, \I{B274}(1986)285.}

\lref\rConnect{P.S.~Green and T.~H\"ubsch: \PRL{61}(1988)1163--1167,
       \CMP{119}(1989)431--441.}

\lref\rCGMS{B.~Greene, D.~Morrison and A.~Strominger:
       \NP{B451}(1995)109--120.}

\lref\rGP{B.R.~Greene and M.~R.~Plesser: \NP{B338}(1990)15-37.}

\lref\rSingS{T.~H\"ubsch: \MPL{A6}(1991)207--216.}

\lref\rBeast{T.~H\"ubsch: {\it \CY\ Manifolds---A Bestiary for
       Physicists} (World Scientific, Singapore, 1992; 2nd ed. 1994).}

\lref\rSL{T.~H\"ubsch and S.-T.~Yau: \MPL{A7}(1992)3277--3289.}

\lref\rKirwan{F.~Kirwan: {\it An Introduction to Intersection Homology
       Theory} (Longman Scientific \&\ Technical, Essex, 1988).}

\lref\rDavid{D.~Morrison: private communication.}

\lref\rReidSing{M.~Reid:  {\it Proc.\,Symp.\,Pure Math.\,\bf46}(1987)345.}

\lref\rCoAS{A.~Strominger: \NP{B451}(1995)96--108.}

\lref\rSuSyM{E.~Witten: {\it J.\,Diff.\,Geom.\,\bf17}(1982)661--692.}

\lref\rMirr{S.-T.~Yau, ed., {\it Essays on Mirror Symmetry},
       (Intl.\ Press, Hong Kong, 1992).}

 %
 %
\newsec{Introduction and Results}\noindent
The close relationship between the cohomology of a (Riemannian) space $X$ and
the Hilbert space of a supersymmetric $\s$-model with target space $X$ has
been studied and known for well over a decade~\rSuSyM. In addition,
Ref.~\rSuSyM\ considered the case of a real, Riemannian manifold $X$ as the
target space for an $N=1$ supersymmetric model. The correspondence derives
from the formal isomorphism between the exterior derivative algebra
\eqn\eExtD{ \big\{\, \rd \,,\, \rd^{\dag} \,\big\} ~=~ \triangle~, }
and the supersymmetry algebra
\eqn\eSuSy{ \big\{\, Q \,,\, \Qb \,\big\} ~=~ H~, }
and the resulting formal isomorphism between the associated complexes.

With more than one supersymmetry, and on complex manifolds, the above
algebras are modified. While the crucial supersymmetry relation~\eSuSy\
remains real
\eqn\eSUSY{ \big\{\, Q_\pm \,,\, \Qb_\pm \,\big\} ~=~ (H\pm p)~, }
the exterior derivative relation become {\it holomorphic}
\eqn\eDolb{ \big\{\, \vd \,,\, \vd^{\dag} \,\big\} ~=~ \triangle_{\vd}
            \qquad\hbox{and}\qquad
            \big\{\, \db \,,\, \db^{\dag} \,\big\} ~=~ \triangle_{\bar\vd}~. }
Nevertheless, the very high degree of analogy between the two basic relations
will ensure the complex generalizations of the results of Ref.~\rSuSyM\ in the
real case. In fact, on K\"ahler manifolds, where
$\triangle_{\vd}=\triangle_{\bar\vd}=\inv2\triangle_{\rd}$,
we can define $\rd_{\pm}\define\vd\pm\db$, whereupon
\eqn\eDolb{ \big\{\, \rd_{\pm} \,,\, \rd_{\pm}^{\dag} \,\big\}
            ~=~ \triangle_{\rd}~. }
So, much as the set of zero-modes (a.k.a.\ the kernel) of $\triangle_{\rd}$
correspond to the de~Rham cohomology and the set of zero-modes of
$\triangle_{\bar\vd}$ correspond to the Dolbeault ($\db$-) cohomology, the
zero-modes of $(H\pm p)$ correspond to the $\Qb_\pm$-cohomology\ft{To be
precise, the translationally invariant zero modes of $H$ are elements of both
the
$\Qb_+$- and the $\Qb_-$-cohomology. Therefore, `the $\Qb_\pm$-cohomology'
should really be taken as an abbreviation for `the $\Qb_+$-cohomology $\cap$
$\Qb_-$-cohomology'. In complex geometry and on a K\"ahler manifold, this
then corresponds to the intersection of the usual $\db$-cohomology and its
conjugate `$\vd$-cohomology'.}. Translationally invariant zero-modes
(those annihilated by the linear momentum $p$) are then also annihilated by
the Hamiltonian, $H$, and so have zero energy. Since $\V{H}\geq0$, these
states are the `ground states', i.e., the (supersymmetric) vacua.
Technically, it is often easier to consider a `twisted' sibling model in which
the $\Qb_\pm$ operators have spin 0 and generate a BRST symmetry. Much as the
original supersymmetry, this `twisted' BRST symmetry produces its associated
complex and the resulting cohomology, which is in a formal 1--1 correspondence
with that of the original supersymmetry. Such `twited' models are then used to
obtain (somewhat restricted) results about the original (`untwisted') model,
but also to provide a more direct mathematical interpretation because they
avoid the complications related to involving spinors.

Most of the literature focuses on cases where $X$ is a smooth manifold. As it
turns out however, suitably (and mildly) singular target spaces are indeed of
interest~\refs{\rConnect,\rCCC,\rSingS,\rBeast,\rCoAS,\rCGMS} and the
relationship between the zero-modes of $H$, i.e., the $\Qb_\pm$-cohomology on
one hand and the available (co)homology theories on singular varieties on the
other ought to be revisited.

The mathematics of singular spaces, and in particular the different types of
cohomology on such, is rather well researched but not very well known
generally; the Reader is referred to Ref.~\rKirwan\ for an introduction in
the general theory. All of the various types of cohomology that can be
defined on singular spaces must reproduce the well known results on smooth
spaces. However, on singular spaces, these cohomology groups do differ in
general.

Partly prompted by this general cohomology theory on singular spaces, and
partly by certain assumptions about the supersymmetric $\s$-model, it was
argued~\rBeast\ that the relevant cohomology on `conifolds'\ft{The term
`conifold' was invented in the last of the works in Ref.~\rCCC, where it was
defined rather loosely. The models examined there only contained nodes.
However, a rather larger but still not very well specified class of
singularities ought to be included~\rSingS, and we will understand
`conifolds' to do so. Furthermore, the recent developments~\rCGMS\ indicate
that the category of spaces ought to be generalized to include `stratified
pseudomanifolds'~\refs{\rExo,\rDavid} (for a definition, see Ref.~\rKirwan).}
ought to be the so-called intersection (co)homology. The intersection homology
is known to be dual to the square-integrable cohomology~\rKirwan. Since the
cohomology representatives correspond to supersymmetric vacuum
wave-functions~\rSuSyM, the square-integrability appears to be a natural
condition. In addition, conifolds appear as interfaces in processes of
topology change of Refs.~\refs{\rConnect,\rCCC}, and the intersection
cohomology appeared to be a good choice (see below).

The purpose of this note is to describe and provide a {\it working
definition\/} for a new type of cohomology for singular spaces. This new
cohomology arises from the recent results of Refs.~\refs{\rCoAS,\rCGMS} which
presents a novel characteristic of superstring models with singular target
spaces. Not surprisingly, this new cohomology satisfies the so-called
`K\"ahler package' requirements. This is exactly as with the intersection
(co)homology and the square-integrable cohomology --- which had been one of
the main reasons in Ref.~\rBeast\ for their use in discussing strings on
singular spaces.

However, unlike any other (co)homology theory known to the present author,
this new cohomology is perfectly symmetric with respect to the still
conjectured but increasingly better accepted `mirror map'~\refs{\rGP,\rMirr}.
This we take as the essential bit of evidence that this new cohomology is {\it
the} cohomology of string theory.

\newsec{The Stringy Singular Cohomology}\noindent
Mostly for the sake of simplicity, we will consider the case where the
singular model, $X^\sharp$ is smooth apart from a finite number of isolated
nodes. Given Arnold's classification of singularities~\rArnold\ and their
small resolutions\ft{Unlike blow-ups, small resolutions {\it always\/} leave
the \CY\ condition $c_1=0$ intact.}~\rReidSing, it should not be too
difficult to generalize the present discussion to all isolated ($A,D,E$)
singularities.

\subsec{The conifold transition}\subseclab\ssConiTrans\noindent
We recall from~\refs{\rConnect,\rBeast} (see Ref.~\rReidSing\ for a more
general discussion) the basic process which connects topologically distinct
\CY\ target spaces. We start with a smooth \CY\ manifold $X^\flat_\e$ the
complex structure of which is made to depend on a complex parameter $\e$ in
such a way that for sufficiently small but nonzero $\e$ the space $X^\flat_\e$
is smooth, but $\lim_{\e\to0}X^\flat_\e=X^\sharp$ is singular. Let
$X^\flat_\e$ acquire $n$ nodes as $\e\to0$. The singular model $X^\sharp$ then
admits a {\it small resolution\/}, $\check{X}$, which again is a smooth \CY\
manifold and has a $\IP^1$ in place of every node of $X^\sharp$. The
transition $X^\flat_\e\to X^\sharp\to\check{X}$ is but the simplest type of a
conifold transition; in the others types, $X^\sharp$ has singularities worse
than nodes, and accordingly, $\check{X}$ will be a more complicated small
resolution\ft{Strictly speaking, $\check{X}\to X^\sharp$ must be a crepant
desingularization: a holomorphic isomorphism except at the inverse images
of the singular points of $X^\sharp$, and which contributes nothing to the
canonical divisor. Thus, the whole process involves only \CY\ models.}.

For $\e\ne0$, the homology group $H_3(X^\flat_\e)$ admits a symplectic
decomposition (perhaps depending on $\e$) into $A_i$ and $B^i$ 3-cycles, with
$i=0,\3,(\inv2b_3{-}1)$, such that:
\eqn\eAnB{ A_i\cap A_j = 0~,\qquad B^i\cap B^j=0~,\quad\hbox{and}\quad
           A_i\cap B^j=\d^j_i~. }

Locally, as $\e\to0$, each node is seen to arise as an $S^3$-shaped 3-cycle
(one of the $A_i$'s) that shrinks to the singular node when $\e=0$. While
$\e\ne0$, each of these $S^3$'s is an element of $H_3(X^\flat_\e)$. However,
not each of the $n$ $S^3$'s need represent an independent homology element
$A_i$. Suppose that there is a single 4-chain (an equivalence class of real
4-dimensional objects), $C_\e$, of which the $n$ $S^3$'s comprise the
boundary. Then a formal sum of the $n$ $S^3$'s (with the signs dictated by
the relative orientation) is homologically trivial, and the last of the $n$
$S^3$'s can be expressed as the negative of the sum of the other $(n{-}1)$
$S^3$'. This leaves $(n{-}1)$ independent $A_i$'s, which we will label
$\hat{A}_i$, the $(n{-}1)$ independent elements of $H_3(X^\flat_\e)$
represented by the $n$ $S^3$'s. Finally, since these $n$ $S^3$'s form
$(n{-}1)$ independent and non-intersecting cohomology elements (a subset of
the $A_i$'s), there must exist $(n{-}1)$ duals to these: $(n{-}1)$ of the
$B^i$'s, which we will label $\hat{B}^i$.

When $\e\to0$, the $n$ `boundary' $S^3$'s shrink to points, whereby
$C_\e\to C_0$ becomes a proper cycle. Thus, in $X^\sharp$, there is one
new 4-cycle $C_0$, and in addition, there remain the $(n{-}1)$ 3-cycles
$\hat{B}^i$. Now, in the intersection homology theory, the $(n{-}1)$
$\hat{B}^i$'s are not counted as proper non-trivial homology elements. In the
square-integrable cohomology theory, one can show that the corresponding forms
are no longer square-integrable.

While somewhat deceiving because of low dimension, the example of a pinched
torus is perhaps instructive. When a torus is deformed so that one of the
`small circles' shrinks to a point, all small circles become homologous to
this point. This point is a singularity of the pinched torus. Formally, the
(middle-dimensional) intersection homology is determined by excising the
singular points, which now turns the $(n{-}1)$ 3-cycles $\hat{B}^i$ into
chains with boundaries, whereupon these become trivial in homology. This is
how intersection homology loses both the $(n{-}1)$ 3-cycles $\hat{A}_i$
represented by the $n$ $S^3$'s that have shrunk to the singular points, and
also their $(n{-}1)$ dual 3-cycles $\hat{B}^i$. In the square-integrable
cohomology, the forms supported on $\hat{B}^i$ become divergent in the limit
when the node forms. In the pinched torus case, this is clear on noting
that the pinching of a cylindrical section is equivalent to stretching it
thinner and longer.

Among 4-cycles, there is the newly formed $C_0$. In the intersection
homology it acquires a dual 2-cycle---traced to the limiting boundaries of
the $(n{-}1)$ chains created from the 3-cycles $C^3$ upon the excision of the
$n$ singular points. These limiting objects however do not exist in the
usual homology where the singular points are not excised. This is how
intersection homology of $X^\sharp$ gains the one dual pair of 4- and
2-cycles while losing $2(n{-}1)$ 3-cycles, which makes it formally isomorphic
to the homology of the small resolution $\check{X}$. In a sense, this is the
reason for the asymmetry of the intersection homology as will be noted more
clearly below.

Finally $\check{X}$, a small resolution of $X^\sharp$, consists of replacing
the $n$ nodes with $S^2$'s, all of which meet the $C_0$ in a single point and
are therefore all dual to that one 4-cycle $C_0$. Thus, the $n$ $S^2$'s are
all homologous and represent a single 2-cycle. This is how small resolution
provides a 2-cycle in the usual homology in place of the formal 2-cycle in the
intersection homology of the singular space. At the same time, each of the $n$
$S^2$'s opens a single `hole' in the $(n{-}1)$ $\hat{B}^i$'s, which thereby
become chains and are no longer non-trivial homology elements. Again, this is
how small resolution loses the 3-cycles remaining after the $n$ $S^3$'s have
shrunk to points, and which intersection homology formally discards.

This leaves none of the $(n{-}1)$ $\hat{A}_i$'s and $\hat{B}^i$'s from
$X^\flat$ surviving into $\check{X}$. On the other hand, the shrinking of the
$n$ $S^3$'s has made $C_0$ into a proper 4-cycle and the process of small
resolution of the nodes has created one dual 2-cycle (homology class) of the
$S^2$'s.

Clearly, had we started with $m$ relations among the $n$ $S^3$, there would
have been $m$ $\hat{C}^a_0$'s and the $n$ $S^2$'s would have formed $m$
homology classes $\hat{D}_a$ in $\check{X}$. This more general situation is
then presented in Table~1 (see also Ref.~\rBeast).

\bigskip
\vbox{
$$\vbox{\offinterlineskip
\hrule height.6pt
\halign{&\vrule width.6pt#&\strut\quad\hfil#\hfil\quad&
                   \vrule#&\strut\quad\hfil#\hfil\quad&
                   \vrule#&\strut\quad\hfil#\hfil\quad&
                   \vrule#&\strut\quad\hfil#\hfil\quad&
                   \vrule#&\strut\quad\hfil#\hfil\quad&\vrule width.6pt#\cr
height2pt&\omit&&\omit&&\omit&&\omit&&\omit&\cr
&$k$& &$\dim'H_k(X^\flat)$&
      &$\dim'H_k(X^\sharp)$&
      &$\dim'I\!H_k(X^\sharp)$&
      &$\dim'H_k(\check{X})$&\cr
height2pt&\omit&&\omit&&\omit&&\omit&&\omit&\cr
\noalign{\hrule height.6pt\vskip1pt\hrule height.6pt}
height2pt&\omit&&\omit&&\omit&&\omit&&\omit&\cr
&$2$& &$0$& &$0$& &$m$& &$m$&\cr
height2pt&\omit&&\omit&&\omit&&\omit&&\omit&\cr
\noalign{\hrule}
height2pt&\omit&&\omit&&\omit&&\omit&&\omit&\cr
&$3$& &$2(n{-}m)$& &$(n{-}m)$& &$0$& &$0$&\cr
height2pt&\omit&&\omit&&\omit&&\omit&&\omit&\cr
\noalign{\hrule}
height2pt&\omit&&\omit&&\omit&&\omit&&\omit&\cr
&$4$& &$0$& &$m$& &$m$& &$m$&\cr
height2pt&\omit&&\omit&&\omit&&\omit&&\omit&\cr} \hrule height.6pt}$$
\vskip0pt
\noindent
{\bf Table 1}: A chart of the varying contributions to the `standard' homology
and intersection homology ($I\!H_*$) for $X^\flat$, $X^\sharp$ and $\check{X}$.}
\bigskip

It should be obvious that $H_k(X^\sharp)$ (or its dual cohomology) cannot
possibly admit a Hodge decomposition, since the number of 3-cycles may well
be odd. Neither can $H_k(X^\sharp)$ (or its dual cohomology) admit Poincar\'e
duality since the $m$ $\hat{C}^a_0$'s in $H_4$ have no duals in $H_2$. Both of
Hodge decomposition and Poincar\'e duality are crucially related to well known
properties of the corresponding $\s$-models\ft{Hodge decomposition is related
to the existence of both left- and right-moving fermions and the associated
left-right symmetry of the 2-dimensional $(2,2)$-supersymmetric field theory
of strings. In $(0,2)$-models which have no left-right symmetry, the relevant
cohomology is valued in a vector bundle (or sheaf) other than the (co)tangent
bundle. Poincar\'e duality corresponds to CPT conjugation which is a symmetry
in the underlying 2-dimensional relativistic field theory.} and so must exist
in any candidate cohomology. The intersection homology certainly does admit
both Hodge decomposition and Poincar\'e duality~\rKirwan, and so is a viable
candidate for a cohomology describing the supersymmetric vacua in a
corresponding supersymmetric
$\s$-model. (These then correspond to massless fields in the effective
spacetime field theory.)

Note that $I\!H_k(X^\sharp)$ is identical to $H_k(\check{X})$ while being
very different from $H_k(X^\flat)$---this is the asymmetry mentioned above.

\subsec{Massless black holes and homology}\noindent
The conifold transition $X^\flat\to X^\sharp\to \check{X}$ is still not fully
understood as a physical phase transition between target spaces for the
heterotic string. However, Type~II superstrings have twice as much spacetime
supersymmetry, which enabled the uncovering of a physical mechanism that
achieves a trouble-free interpolation between the target spaces $X^\flat$,
$X^\sharp$ and $\check{X}$~\refs{\rCoAS,\rCGMS}. The results are merely retold
in the sequel; the Reader is referred to Ref.~\refs{\rCoAS,\rCGMS} for further
details of the physics mechanism in this phase transition.

The $N=2$ spacetime supersymmetry of Type~II strings permits the existence of
(BPS) states which preserve half of the supersymmetry, and the mass of which
is bounded (and in the extremal case equal) to their charge. Such states
accompany the not-yet-shrunk $(n{-}m)$ $\hat{A}_i$'s of $X^\flat_\e$ and their
mass is proportional to the volume of the $S^3$'s representing these cycles.
The effective field theory of the Type~IIB strings contains a 4-form
potential. Field configurations where the field strength (5-form) $F$
has a non-zero integral over the 3-cycle $\hat{A}_i$ within the \CY\ 3-fold
appear, in the complementary 4-dimensional Minkowski space, as monopoles
with the field strength (2-form) $\int_{\hat{A}_i}F$.

As the $S^3$'s shrink to points in the $\e\to0$ limit, these states become
massless. So, the effective field theory derived from a Type~II string
theory constructed on $X^\flat_\e$ loses $(n{-}m)$ `elementary' massless
states as $\e\to0$, but promptly gains their replacements in the form of these
by now massless BPS states. Thus, the low energy effective physics counts
$2(n{-}m)$ massless states throughout the variation of $\e$, from nonzero
values to {\it and including\/} $\e=0$.

Furthermore, there also exists $m$ states corresponding to the $m$ relations
among the $n$ nodes, and these states essentially provide a `mirror story'.
They are massive and have nonzero vacuum expectation values in the model built
upon $\check{X}$, but become massless as the $n$ $S^2$'s are shrunk to points.
Recall that these $n$ $S^2$'s form $m$ 2-cycles, $\hat{D}_a$, which are dual
to the 4-cycles $\hat{C}^a_0$, as described above. The $m$ BPS states then
can be seen to stem from field configurations of the 3-form potential in the
Type~IIA theory, the field strength $G$ of which have non-zero integrals
$\int_{\hat{D}_a}G$, and which then appear as field strength 2-forms of
monopoles (one for each $\hat{D}_a$) in the 4-dimensional Minkowski spacetime.
Again, each of these $m$ states has a mass proportional to the area of the
$S^2$'s representing its 2-cycle.

Thus, in the conifold transition, the varying set of massless fields---with
these massless BPS states included!---is in a sense complementing the
situation presented in Table~1, and the results are summarized in Table~2.

\bigskip
\vbox{
$$\vbox{\offinterlineskip
\hrule height.6pt
\halign{&\vrule width.6pt#&\strut\quad\hfil#\hfil\quad&
                   \vrule#&\strut\quad\hfil#\hfil\quad&
                   \vrule#&\strut\quad\hfil#\hfil\quad&
                   \vrule#&\strut\quad\hfil#\hfil\quad&
                   \vrule#&\strut\quad\hfil#\hfil\quad&\vrule width.6pt#\cr
height2pt&\omit&&\omit&&\omit&&\omit&&\omit&\cr
&$k$& &$\dim'H^k(X^\flat)$&
      &$\dim'H^k(X^\sharp)$&
      &$\dim'S\!H^k(X^\sharp)$&
      &$\dim'H^k(\check{X})$&\cr
height2pt&\omit&&\omit&&\omit&&\omit&&\omit&\cr
\noalign{\hrule height.6pt\vskip1pt\hrule height.6pt}
height2pt&\omit&&\omit&&\omit&&\omit&&\omit&\cr
&$2$& &$0$& &$0$& &$m$& &$m$&\cr
height2pt&\omit&&\omit&&\omit&&\omit&&\omit&\cr
\noalign{\hrule}
height2pt&\omit&&\omit&&\omit&&\omit&&\omit&\cr
&$3$& &$2(n{-}m)$& &$(n{-}m)$& &$2(n{-}m)$& &$0$&\cr
height2pt&\omit&&\omit&&\omit&&\omit&&\omit&\cr
\noalign{\hrule}
height2pt&\omit&&\omit&&\omit&&\omit&&\omit&\cr
&$4$& &$0$& &$m$& &$m$& &$m$&\cr
height2pt&\omit&&\omit&&\omit&&\omit&&\omit&\cr} \hrule height.6pt}$$
\vskip0pt
\noindent
{\bf Table 2}: A table of the varying contributions to the massless field
spectrum in the effective field theory of Type~II superstrings built upon
$X^\flat$, $X^\sharp$ and $\check{X}$. The states are listed as
corresponding to cohomology groups.}
\bigskip

Notice that rather than being equal either $H^*(X^\flat)$ or $H^*(\check{X})$,
this set of new cohomology groups $S\!H^*(X^\sharp)$ are in fact,
formally, the {\it union\/} of $H^*(X^\flat)$ and $H^*(\check{X})$! It is
precisely these $S\!H^*(X^\sharp)$ that are herein proposed as the
(super)stringy singular cohomology\ft{The acronym SSC, albeit naturally
formed, is perhaps best avoided for being ominous in view of the tightness of
the present funding situation.}.

\subsec{Calculating the (super)stringy singular (co)homology}\noindent
No general cohomology theory for singular spaces that would produce this new
cohomology group is know as yet to the present author. However, modeled on the
Corollary on p.~313 of Ref.~\rIHom, a {\it working definition\/} of the
(super)stringy singular homology, $S\!H_*(X^\sharp)$ is provided below. The
corresponding cohomology, $S\!H^*(X^\sharp)$, is then defined as the formal
dual. Some rough characteristics of the elements of $S\!H^*(X^\sharp)$ by
which it differs from (the square-integrable) $H^*_{(2)}(X^\sharp)$ will also
transpire.

For practical purposes of calculating the intersection homology of the
singular limit $X^\sharp = \lim_{\e\to0}X^\flat_\e$, we quote the Corollary
from Ref.~\rIHom\ (see also \SS~7.3 of Ref.~\rBeast):
\bigskip
\vbox{\narrower\noindent
{\bf Corollary}\\
Let $X$ be an $n$-fold with a single isolated singularity, $x$. Then
$$ I\!H_k(X) =
 \cases{H_k(X) & $k>n$~,\cr
         \noalign{\vglue1mm}
        {\rm Im}\big[\, H_n(X{-}x) \to H_n(X)\big] & $k=n$~,\cr
         \noalign{\vglue1mm}
        H_k(X{-}x) & $k<n$~.\cr}
$$}
\noindent
For practical purposes of calculating the (super)string singular homology of
the singular limit $X^\sharp = \lim_{\e\to0}X^\flat_\e$, we then propose
\bigskip
{\bf Definition}\\ \indent
Let $X$ be an $n$-fold with a single isolated singularity, $x$. Then
\eqn\eDef{ S\!H_k(X) =
 \cases{H_k(X) & $k>n$~,\cr
         \noalign{\vglue1mm}
        H_n(X{-}x) \cup H_n(X) & $k=n$~,\cr
         \noalign{\vglue1mm}
        H_k(X{-}x) & $k<n$~.\cr} }
\noindent
The definition generalizes to cases where the singular locus, $x$, consists of
more than one point and in fact of several component subspaces of $X$---just
as it does in the case of the above corollary for $I\!H_k$. Also, the formal
union in~\eDef\ may be better understood as an extension of $H_n(X)$ by
$\ker\big[H_n(X{-}x)\to H_n(X)\big]$, or perhaps of $H_n(X{-}x)$ by
coker$\big[H_n(X{-}x)\to H_n(X)\big]$. The precise nature of this extension is
to be determined from the as yet unspecified general cohomology theory.
\bigskip

We note that both the above corollary and the above {\it working definition\/}
provide a prescription for calculating $I\!H_k$ and $S\!H_k$, and in terms
intrinsic to the singular space itself and independent of any possible
smoothings (through deformation, or any resolution). We therefore {\it
expect\/} that a well-defined and fully intrinsic general homology theory for
$S\!H_k$ can also be found, just as there is one for $I\!H_k$~\rKirwan. This
task is however beyond our present scope.

The situation with the ring structure of $S\!H_*$ is somewhat different.
Since dual pairs $\hat{A}_i,\hat{B}^j$ in the middle dimensional homology
$S\!H_n(X^\sharp)$ come from different spaces in the definition,
$H_n(X{-}x)$ and $H_n(X)$ respectively, their intersection ring structure is
not as well defined as for the rest of the $n$-cycles which pass through the
conifold transition unaffected. It seems however reasonable to consider the
ring structure of $H_n(X^\flat_\e)$ as a function of $\e$, and assign the
limit $\e\to0$ of this ring structure to $S\!H_n(X^\sharp)$. Similarly, the
part of $S\!H_k(X^\sharp)$ with $k>n$ or $k<n$, represented by the
$\hat{C}^a_0,\hat{D}_a$ and which is absent from $H_k(X^\flat_\e)$, is equally
difficult to deal with. These cohomology elements do have their counterparts
in $H_k(\check{X})$ and a well-defined ring structure there---both the
classical (generated from the wedge product) and also the quantum one
(deformed as a function of the family of rational curves in $\check{X}$). Let
$\j$ denote the root-mean-square area of the $S^2$ of $\check{X}$ which
replaced the nodes of $X^\sharp$. Then, the limit $\j\to0$ of the ring
structure of $H_k(\check{X}_\j)$, for $k\neq0$, may be assigned as the ring
structure of $S\!H_k(X^\sharp)$.

In other words, each \CY\ space is fibred over its respective space of
complex structures and extended and complexified K\"ahler class (the latter
defined more precisely in Ref.~\rExo). The smooth spaces $X^\flat_\e$ and
$\check{X}_\j$ each occur in such a family (although only the parameters
relevant for the conifold transitions are noted explicitly). In either of these
families the (quantum) ring structures of $H^{3-*,*}$ and of $H^{*,*}$ depend
on the moduli parameters. The singular space(s) $X^\sharp$ appear on the
joining interface between the $X^\flat_\e$ family and the $\check{X}_\j$
family, and so acquire a cohomology ring structure in the limit.

These assignments obviously depend on the choice(s) of the smoothing
deformation $X^\sharp\to X^\flat_\e$ and the choice(s) of resolution of
singularities $X^\sharp\to\check{X}_\j$. This dependence on
deformation i.e.\ resolution is perhaps undesirable as it is not intrinsic to
the singular space $X^\sharp$ itself. However it seems to make perfect sense
when considering the connected families of spaces
$X^\flat_\e,X^\sharp,\check{X}_\j$ fibered over the joined total moduli space
(both the space of complex structures and the extended and complexified
K\"ahler classes~\refs{\rConnect,\rCCC,\rBeast,\rAGM}).

\newsec{Some Notable Properties}\noindent
Besides being simply another set of (co)homology groups defined on (certain)
singular spaces, $S\!H_*$ and $S\!H^*$ exhibit a few properties by which they
are distinguished from the rest. We now turn to examine two: relation to mirror
symmetry and the properties commonly grouped under the name `K\"ahler
package'.

\subsec{Mirror symmetry}\noindent
Consider a conifold transition $M^\flat_\e\to M^\sharp\to\check{M}_\j$ and
recall that in this process $b_3$ decreases while $b_2{+}b_4$ increases.
Consider now the corresponding families of mirror spaces
$W^\flat_\e,W^\sharp,\check{W}_\j$. The mirror map exchanges (the
r\^oles in the cohomology valued in) the tangent and the cotangent bundle,
whereby $b_3$ and $b_2{+}b_4$ are exchanged. Therefore, in the mirror of a
conifold transition, the Betti numbers change in the opposite direction.
Assuming that the spaces $W^\flat_\e,W^\sharp,\check{W}_\j$ are also connected
through a conifold transition, we then have a mirror pair of conifold
transitions:
\eqna\eXXX
 $$\eqalignnotwo{
 M^\flat_\e   &\to ~M^\sharp &\to ~\check{M}_\j~, &\eXXX{a}\cr
 \check{W}_\j &\to ~W^\sharp &\to ~W^\flat_\e~.   &\eXXX{b}\cr}$$
Consider now using $I\!H_k$ for all of these spaces (in the smooth cases
$I\!H_k=H_k$, by definition). By virtue of the results in Table~1, we obtain
the rather asymmetric pair:
\eqna\eXXX
 $$\eqalignnotwo{
 I\!H_k(M^\flat_\e)   &\too{\neq} ~I\!H_k(M)^\sharp
                      &\too{\approx} ~I\!H_k(\check{M}_\j)~, &\eXXX{a}\cr
 I\!H_k(\check{W}_\j) &\too{\approx} ~I\!H_k(W)^\sharp
                      &\too{\neq} ~I\!H_k(W^\flat_\e)~.   &\eXXX{b}\cr}$$
By contrast, using the results in Table~2, $S\!H_k$ produces a perfectly
mirror-symmetric pair of conifold transitions:
\eqna\eXXX
 $$\eqalignnotwo{
 S\!H_k(M^\flat_\e)   &\too{\subset} ~S\!H_k(M)^\sharp
                      &\too{\supset} ~S\!H_k(\check{M}_\j)~, &\eXXX{a}\cr
 S\!H_k(\check{W}_\j) &\too{\subset} ~S\!H_k(W)^\sharp
                      &\too{\supset} ~S\!H_k(W^\flat_\e)~. &\eXXX{b}\cr}$$
For the smooth spaces $S\!H_k=H_k$, by definition.

Also, $S\!H_k(X^\sharp)$ is bigger than either $H_k(X^\flat)$ or
$H_k(\check{X})$ --- in good agreement with the physics intuition that
string theory on $X^\sharp$ ought to possess extra zero-modes, stemming from
extra symmetries (continuous or discrete) and forced by anomaly cancellation
of these symmetries. This general expectation is indeed borne out in the
analysis of Type~II strings~\refs{\rCoAS,\rCGMS}, the analysis of which was
facilitated by the $N=2$ supersymmetry. In the heterotic string case, the
$N=1$ supersymmetry does not provide enough rigidity to derive analogous
results, and in addition, the heterotic string lacks the 4- and 3-forms of
the Type~II strings the topologically non-trivial configurations of which
provide for the BPS (monopole-like) states. As stated in Ref.~\rCoAS, the
physical mechanism of the conifold transition remains a mystery for the
heterotic strings.

\subsec{The K\"ahler package}\noindent
One of the major arguments in favor or using $I\!H_k$ (and its formal dual
$I\!H^k$) for singular spaces in Ref.~\rBeast\ was the fact that this
(co)homology features the properties collectively referred to as `the
K\"ahler package':
\item{1.} Hodge decomposition:
\eqn\eXXX{ H^r(X) = \bigoplus_{p+q=r}H^{p,q}(X)~; }
\item{2.} Complex conjugation:
\eqn\eXXX{ \overline{H^{p,q}(X)} = H^{q,p}(X)~; }
\item{3.} Poincar\'e duality:
\eqn\eXXX{ H^{p,q}(X) \buildrel*\over\approx H^{n-q,n-p}(X)~; }
\item{4.} K\"unneth formula:
\eqn\eXXX{ H^{r,s}(X{\times}Y) = \bigoplus_{p+p'=r\atop q+q'=s}
          H^{p,q}(X){\otimes}H^{p',q'}(Y)~. }
\item{5.} Lefschetz $SL(2,\IC)$ action (with $\w$ the K\"ahler form on the
$n$-fold $X$):
\eqn\eLef{{\eqalign{
   & L(\h)\define \w\wedge\h~,\qquad \L\define ({-})^n*L*~,\qquad
   \forall\h\in H^{*,*}(X)~,\cr
   & h\define\big[L\,,\,\L\big]~,\qquad
     \big[h\,,\,L\big] = L~,\qquad \big[h\,,\,\L\big] = -\L~,\cr}}}
and the induced decomposition into irreducible $SL(2,\IC)$ representations
(the so-called `Lefschetz Hard Theorem').

The first four parts of the K\"ahler package have immediate counterparts in
the 2-dimensional field theory underlying the string propagation on $X$.
These are easily seen to be featured by both $I\!H_k$ and by $S\!H_k$.

On \CY\ spaces, the last property may literally be doubled~\rSL\ owing to
mirror symmetry. That is, {\it there really exist two `orthogonal' Lefschetz
$SL(2,\IC)$ actions\/}. The conventional one acts `vertically', whereas the
second one acts `horizontally' in the Hodge diamond. While the conventional
one is generated by wedge products with the K\"ahler form~\eLef, the second
$SL(2,\IC)$ action is the mirror-preimage of the conventional Lefschetz
$SL(2,\IC)$ action on the cohomology of the mirror space. Both $SL(2,\IC)$
actions induce a decomposition into irreducible $SL(2,\IC)$ representations.
Finally, at least for \CY\ 3-folds, these two $SL(2,\IC)$ actions commute, and
rather trivially, since the cohomology elements not annihilated by one action
are annihilated by the other. While the physics application and implication of
these actions is not yet clear, requiring their existence poses rather severe
restrictions on the cohomology groups, and cohomology theory in general.

Just as with the discussion of the ring structure developed on $S\!H_k$, we
again define the `vertical' and `horizontal' Lefschetz $SL(2,\IC)$ action on
$X^\sharp$ as a limit from $\check{X}_\j$ and from $X^\flat_\e$,
respectively. While not satisfactory as an intrinsic definition, this
certainly makes perfect sense for the connected families of
$X^\flat_\e,X^\sharp,\check{X}_\j$.

In fact, already with $I\!H^k(X^\sharp)$, one can trace the conventional
(vertical) $SL(2,\IC)$ action as the limit of the same on $H^k(\check{X}_\j)$,
to which it is isomorphic (for any $\j$) as a vector space. Comparing tables 1
and 2, we see that, on an $n$-fold, $I\!H_*\subseteq S\!H_*$ and
$I\!H_k\subset S\!H_k$ only for $k=n$; the analogous is true for the formal
duals $I\!H^k$ and $S\!H^k$. So, except for the middle dimension, the proof
of the conventional Lefschetz $SL(2,\IC)$ action for $I\!H^k$~\rKirwan\ will
also apply for $S\!H^k$. On \CY\ 3-folds---the case of our immediate
interest---all middle dimensional cohomology is annihilated by the raising and
lowering operators of $SL(2,\IC)$ simply because the $(3{-}q{\pm}1,q{\pm}1)$
cohomology is empty. This then obviously holds both for $I\!H^3(X^\sharp)$ and
$S\!H^3(X^\sharp)$. The argument for the `horizontal' $SL(2,\IC)$ action is
precisely analogous, upon exchanging the r\^oles of the rows and columns of
the Hodge diamond, and the limit obtained from
$X^\flat_\e$ by letting $\e\to0$.

Both of the $SL(2,\IC)$ actions are expected to degenerate partly in the
singular limit $\e,\j\to0$, since non-degeneracy is guaranteed only for
generic models~\rSL. At the location of such degeneration, some of the the
irreducible $SL(2,\IC)$ representations decompose into smaller ones. For
example, the isomorphism
$\w_\j:H^{1,1}(\check{X}_\j)\too{\sim}H^{2,2}(\check{X}_\j)$ for a
family of 3-folds\ft{Note: $\j$ parametrizes the choice of the K\"ahler form,
not the complex structure. However, we fiber the \CY\ spaces over the combined
space of complex structures and complexified and extended K\"ahler classes as
the moduli space.} is realized by wedging with the $\j$-dependent K\"ahler form
$\w_\j$. It pairs a $(1,1)$-form with a $(2,2)$-form into an $SL(2,\IC)$
doublet. At special values $\j'$, this map will develop a non-zero kernel and
cokernel. Some of the $(1,1)$-forms that are now annihilated by $\w_{\j'}$,
and a corresponding number of $(2,2)$-forms are no longer obtainable as
$\w_{\j'}{\wedge}\h$ for $\h\in H^{1,1}(\check{X}_{\j'})$. These unpaired
$(1,1)$- and $(2,2)$-forms have now become pairs of 1-dimensional (trivial)
representations of $SL(2,\IC)$. Ref.~\rSL\ gives concrete examples for the
degeneration the `horizontal' Lefschetz $SL(2,\IC)$ action, but the same
applies quite clearly for the standard (`vertical') action also.

In a way, the (co)homology $S\!H_*$ ($S\!H^*$) may be considered the answer
to the question if there is a (co)homology on (at least some) singular spaces,
and other than $I\!H_*$ ($I\!H^*$ and the $L^2$-cohomology), that features the
K\"ahler package.

\subsec{Some open considerations}\noindent
Besides the Hodge decomposition of the cohomology (the first point in the
K\"ahler package), the $\Qb_\pm$-cohomology of the 2-dimensional $\s$-model
also exhibits the Hodge decomposition of forms. This amounts to the statement
that any cohomology element is represented by a harmonic form (annihilated by
$H{\pm}p$) up to the addition of a $\Qb_\pm$-exact and a $Q_\pm$-exact term.
It would be interesting to see if this property extends, perhaps with respect
to the $\rd_\pm$ operators defined for Eq.~\eDolb, to $S\!H^*$.

The definition~\eDef\ may well appear to be rather {\it ad hoc\/} and it is
tempting to speculate if it could be derived from the underlying structure of
string theories. The two natural approaches would involve either the
(canonical, Hamiltonian) configuration space of string theory, the loop space
${\cal L}X$, or the (relativistically covariant, Lagrangian) configuration
space, the space of maps $\S\to X$, where $\S$ is the universal curve.

At least in the $n=1$ case (tori), the loop-space approach suggests a physical
reason for retaining the 1-cycles represented by the `small' circles even
after one of them has been shrunk to a point. Possessing a non-zero tension,
the string simply cannot be made to shrink to the singular point. This, of
course is also consistent with the $R\iff{1\over R}$ duality --- one of the
first duality relations found in string theory. The situation in $n>1$ is
then likely to be more involved and will perhaps relate to some of the more
recently uncovered duality relations.

Just as the relation to mirror symmetry seems to be a strong indication in
favor of $S\!H_k$, the latter approach introduces another possibility to
relate $S\!H_k$ to at least some of the newly discovered dualities of string
theory. To that end, note that maps $\S_g\to X$, with $\S_g$ a Riemann surface
of genus $g$, may be thought of as graphs in the space $\S_g\times X$, whence
string theory may be understood in terms of a field theory in the
12-dimensional $\S_g\times X$ constrained however to the graphs of $\S_g\to
X$. In turn, it is becoming clearer that the `master theory' for the dualities
is probably a 12-dimensional theory the lower dimensional limits of which
exhibit local supergravity. Yet, this 12-dimensional theory (`F-theory')
itself cannot be a supergravity theory since the (smallest) spinor
representation is too large to be mapped, 1--1, to tensorial representations of
(integral) spin 2 and less. Therefore, this 12-dimensional theory must be a
constrained $N=\inv2$ theory, and may well be describing the graph of the
string in the 12-dimensional space $\S_g{\times}X$. In this constrained
12-dimensional theory and in particular on its cohomology, the duality
involutions should have a simpler action (much as the non-linear M\"obius
transformation becomes linear when acting on the homogeneous coordinates of
$\IP^1$). This could provide additional clues about the cohomology of this
master theory, and so also about the induced cohomology of the target space of
the string propagation --- hopefully, the $S\!H^*$ defined herein or its
generalization.

In closing, a possible mathematical framework in which to define intrinsically
the $S\!H^*$ and $S\!H_*$ groups appears natural to suggest. Recall that the
`stringy cohomology' on Coxeter orbifolds ($T^6/D$, where $D$ is the action of
a finite group) introduced `twisted states': string configurations which are
contractible in conventional geometry, but are obstructed from contracting by
the string dynamics (tension). Such cohomology elements are localized at the
finite quotient singularities of these orbifolds~\rDHVW. In this case, the
mathematical framework for a rigorous definition of this cohomology turned out
to be equivariant $K$-theory~\rAtSe, owing to the fact that such orbifolds are
global quotients by a finite group $G$. The present case is more complicated,
because the singularities now do not stem from a {\it global\/} quotient by the
non-free action of a discrete group $G$. Consequently, there are no
$G$-bundles to define a suitable $K$-theory and a $G$-equivariant cohomology.
Instead, there exist local group actions, restricted to the singular points
and the tangent cones centered at the singular points. Instead of a global
$G$-bundle for a $K$-theory, then, sheaves with only locally supported group
actions seem natural candidates, leading to a suitably generalized,
$S\!K$-theory---hopefully the underlying theory for a rigorous definition of
$S\!H^*$ and $S\!H_*$.

 %
\vfill
\noindent{\bf Acknowledgments}:
Very helpful discussions with David Morrison and Paul Green and gratefully
acknowledged. This work was supported by the Department of Energy grant
DE-FG02-94ER-40854.

\vfill\eject

\listrefs

 %
\bye